\documentclass[10pt,conference]{IEEEtran}
\usepackage{amssymb,amsmath,graphics,graphicx,epsfig}
\usepackage{subfigure}
\usepackage{algorithm}
\usepackage{algorithmic} 
\usepackage{verbatim}
\usepackage{url}
\usepackage{citesort}
\usepackage{balance}

\newcommand{\Aut}[1]{\mathrm{Aut}(\mathcal{#1})}
\newcommand{\Autc}{\mathrm{Aut}(\mathcal{C})}

\newcommand{\TG}{\mathbf{TG}(H)}
\newcommand{\TGm}{\mathbf{TG}(H')}
\newcommand{\C}{\mathcal{C}}
\newcommand{\dmin}[2]{d_{\rm {min}}(\mathcal{#1}#2)}

\begin{document}

\title{Iterative Decoding on Multiple Tanner Graphs Using Random Edge Local Complementation}
\author{ \IEEEauthorblockN{Joakim Grahl Knudsen, Constanza
    Riera\IEEEauthorrefmark{1}, Lars Eirik Danielsen, Matthew G. Parker, 
    and Eirik Rosnes}
  \IEEEauthorblockA{Dept. of Informatics, University of Bergen, Thorm{\o}hlensgt. 55, 5008 Bergen, Norway\\
    Email: \{joakimk, larsed, matthew, eirik\}@ii.uib.no}
  \IEEEauthorblockA{\IEEEauthorrefmark{1}Bergen University College, 
    Nyg{\aa}rdsgt. 112, 5008 Bergen, Norway. Email: csr@hib.no} }

\maketitle
\begin{abstract}
  In this paper, we propose to enhance the performance of the
  sum-product algorithm (SPA) by interleaving SPA iterations with a
  random local graph update rule. This rule is known as edge local
  complementation (ELC), and has the effect of modifying the Tanner
  graph while preserving the code. We have previously shown how the
  ELC operation can be used to implement an iterative permutation
  group decoder (SPA-PD)--one of the most successful iterative
  soft-decision decoding strategies at small blocklengths. In this
  work, we exploit the fact that ELC can also give structurally
  distinct parity-check matrices for the same code. Our aim is to
  describe a simple iterative decoder, running SPA-PD on distinct
  structures, based entirely on random usage of the ELC
  operation. This is called SPA-ELC, and we focus on small blocklength
  codes with strong algebraic structure. In particular, we look at the
  extended Golay code and two extended quadratic residue codes. Both
  error rate performance and average decoding complexity, measured by
  the average total number of messages required in the decoding,
  significantly outperform those of the standard SPA, and compares
  well with SPA-PD. However, in contrast to SPA-PD, which requires a
  global action on the Tanner graph, we obtain a performance
  improvement via local action alone. Such localized algorithms are of
  mathematical interest in their own right, but are also suited to
  parallel/distributed realizations.
%
\end{abstract}

\section{Introduction}
Inspired by the success of iterative decoding of low-density
parity-check (LDPC) codes, originally introduced by Gallager
\cite{Gal62} and later rediscovered in the mid 1990's by MacKay and
Neal \cite{mackMN}, on a wide variety of communication channels, the
idea of iterative, soft-decision decoding has recently been applied to
classical algebraically constructed codes in order to achieve
low-complexity belief propagation decoding
\cite{jia06,jia04,halford_NEW,hehn,beery,knudsen08}. Also, the
classical idea of using the automorphism group of the code, $\Aut{C}$,
to permute the code, $\C$, during decoding (known as
\textit{permutation decoding} (PD) \cite{macw_pd}) has been
successfully modified to enhance the sum-product algorithm (SPA) in
\cite{halford_NEW}. We will denote this algorithm by
SPA-PD. Furthermore, good results have been achieved by running such
algorithms on several structurally distinct representations of $\C$
\cite{jia06,hehn}. Both Reed-Solomon and Bose-Chaudhuri-Hocquenghem
(BCH) codes have been considered in this context. Certain
algebraically constructed codes are known to exhibit large minimum
distance and a non-trivial $\Aut{C}$. However, additional properties
come into play in modern, graph-based coding theory, for instance,
sparsity, girth, and trapping sets \cite{richurbook,ric03}.
Structural weaknesses of graphical codes are inherent to the
particular parity-check matrix, $H$, used to implement $\C$ in the
decoder. This matrix is a non-unique $(n-k)$-dimensional basis for the
null space of $\C$, which, in turn, is a $k$-dimensional
subspace of $\{0,1\}^n$. Although any basis (for the dual code,
$\C^{\bot}$) is a parity-check matrix for $\C$,
their performance in decoders is not uniform. $H$ is said to be in
standard form if the matrix has $n-k$ weight-1 columns.
The weight of $H$ is the number of non-zero entries, and the minimum
weight is lower-bounded by $(n-k)\dmin{C}{^\bot}$, where
$\dmin{C}{^\bot}$ denotes the minimum distance of $\C^{\bot}$.
It is well-known that $H$ can be mapped into a bipartite (Tanner)
graph, $\TG$, which has an edge connecting nodes $v_i$ and $u_j$
\textit{iff} $H_{ji} \ne 0$. Here, $v_i, 0 \le i < n$, refers to the
bit nodes (columns of $H$), and $u_j, 0 \le j < n-k$, refers to the
check nodes (rows of $H$). The local neighborhood of a node, $v$, is
the set of nodes adjacent to $v$, and is denoted by
$\mathcal{N}_v$. The terms standard form and weight extend trivially
to $\TG$. In the following, we use bold face notation for vectors, and
the transpose of $H$ is written $H^T$.

This paper is a continuation of our previous work on edge local
complementation (ELC) and iterative decoding, in which selective use
of ELC (with preprocessing and memory overhead) equals
SPA-PD \cite{knudsen08}. In this work, we use ELC in a truly random,
online fashion, thus simplifying both the description and application
of the proposed decoder. The key difference from our previous work is
that we do not take measures to preserve graph isomorphism, and
explore the benefits of going outside the automorphism group of the
code. This means that we alleviate the preprocessing of suitable ELC
locations (edges), as well as the memory overhead of storing and sampling from
such a set during decoding. Our proposed decoding algorithm can be
thought of as a combination of SPA-PD \cite{halford_NEW} and multiple
bases belief propagation \cite{hehn}. We also discuss the modification
of the powerful technique of damping to a graph-local
perspective.

\section{The ELC Operation}
\begin{figure}[t]
\centering
\subfigure[]{\label{elc1}\includegraphics[scale=0.35]{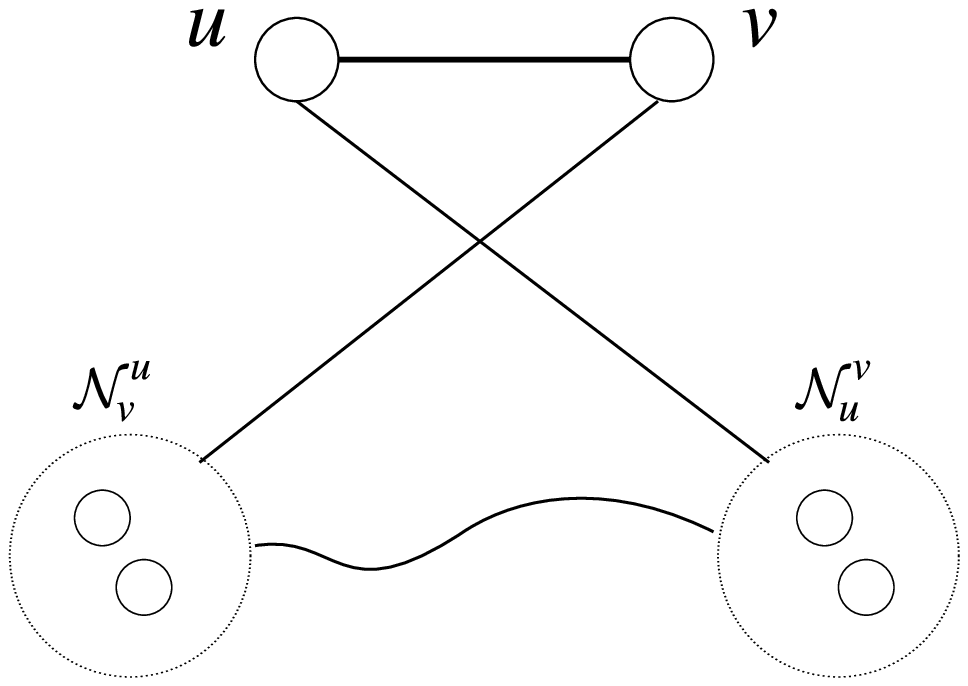}}
\hspace{0.1cm}
\subfigure[]{\label{elc2}\includegraphics[scale=0.35]{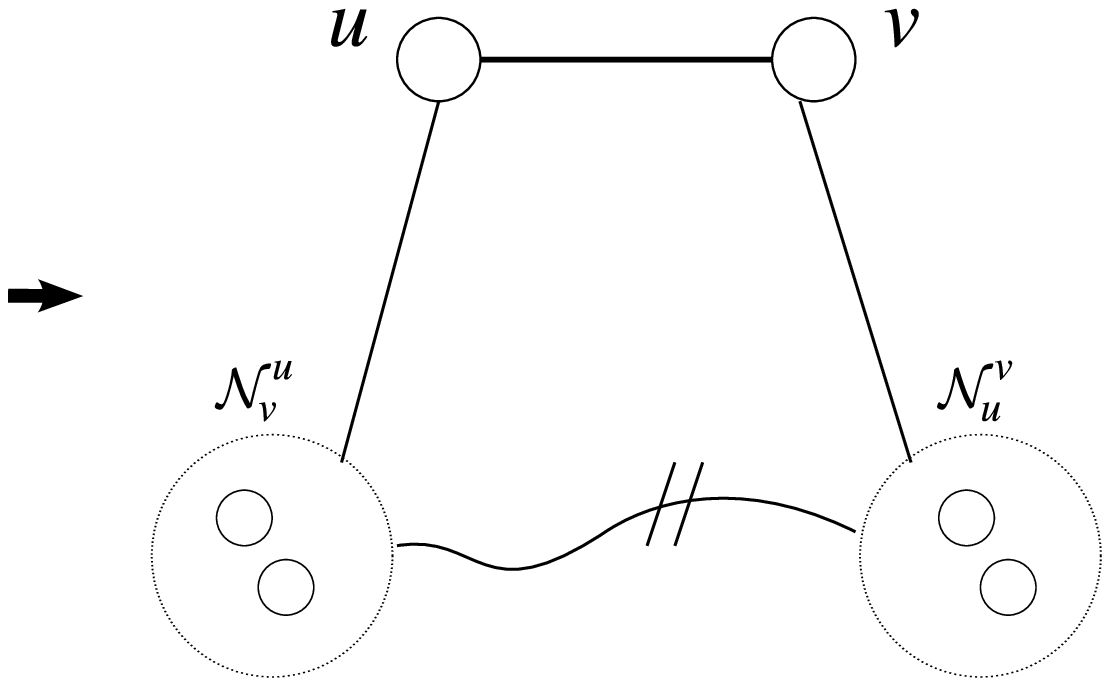}}
\caption{ELC on edge $(u,v)$ of a bipartite graph. Doubly slashed
  links mean that the edges connecting the two sets have been complemented.}
\label{elc_test}
\end{figure}
\begin{figure}[t]
\centering
\subfigure[$G$]{\label{gb}\includegraphics[scale=0.5]{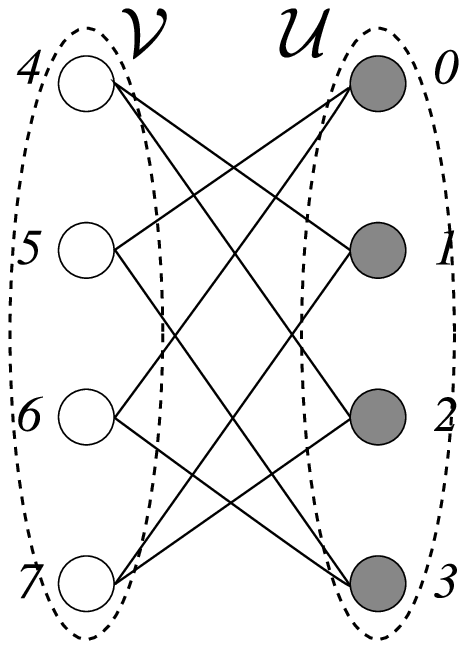}}
\hspace{0.2cm}
\subfigure[$\mathbf{TG}(H)$]{\label{ga}\includegraphics[scale=0.5]{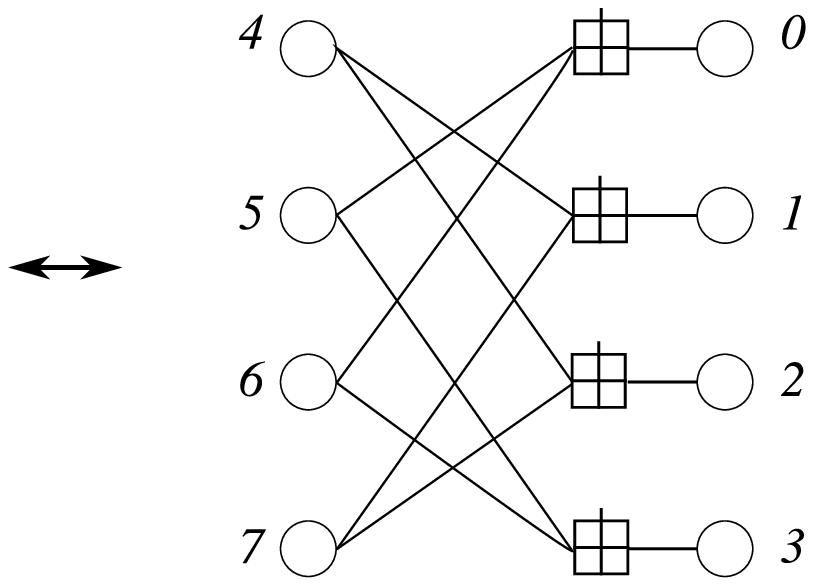}}
\subfigure[$G'$]{\label{ge}\includegraphics[scale=0.5]{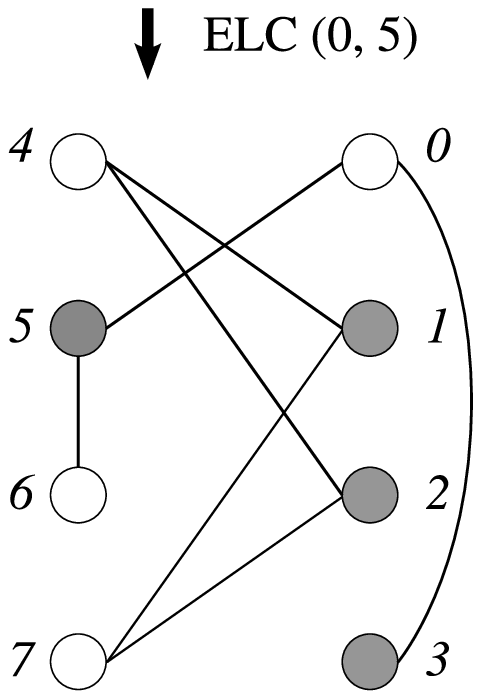}}
\hspace{0.2cm}
\subfigure[$\TGm$]{\label{gd}\includegraphics[scale=0.5]{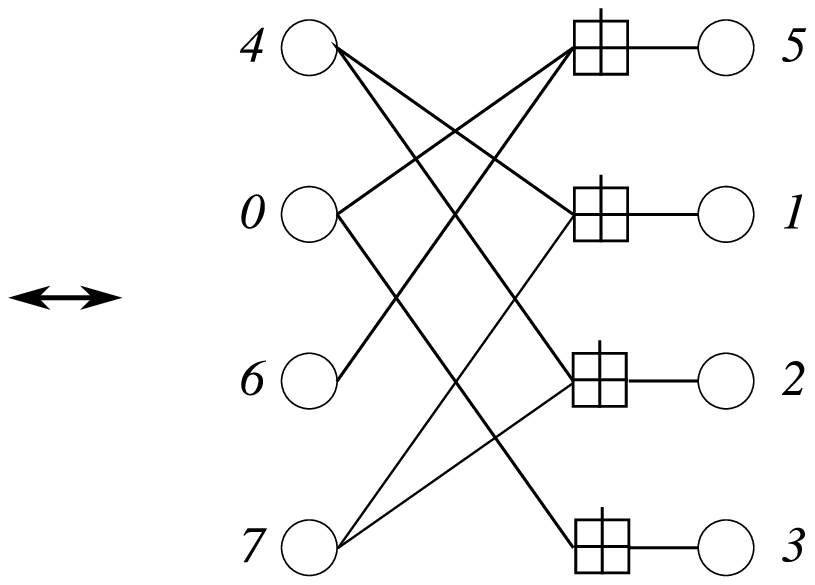}}
\caption{(a) and (c) show ELC on the edge $(0,5)$ of a small simple
  graph $G$. The corresponding Tanner graphs, in (b) and (d), are
  distinct structures (e.g., the weight of $G$ and $G'$ is not the
  same) for the same toy $[8,4,2]$ code. This code has a total of
  three structures in its s-orbit.}
\label{pivot}
\end{figure}

The operation of ELC \cite{bouchet,ELC_danielsen,riera_orbits}, also
known as Pivot, is a local operation on a simple graph (undirected
with no loops), $G$, which has been shown to be useful both for code
equivalence and classification \cite{ELC_danielsen}, and for decoding
purposes \cite{knudsen08}. It has recently been identified as a useful
local unitary primitive to be applied to \textit{graph states}
\cite{riera_orbits}--a proposed paradigm for quantum computation
\cite{EntGraph}. Fig. \ref{elc1} shows $G_{\mathcal{N}_u \cup
  \mathcal{N}_v}$, the local subgraph of a bipartite graph induced by
nodes $u, \ v$, and their disjoint neighborhoods which we denote by
$\mathcal{N}_u^v \triangleq \mathcal{N}_u\setminus\{v\}$ and
$\mathcal{N}_v^u \triangleq \mathcal{N}_v \setminus\{u\}$,
respectively. ELC on a bipartite graph is described as the
complementation of edges between these two sets; $\forall v' \in
\mathcal{N}_u^v$ and $\forall u' \in \mathcal{N}_v^u$, check whether
edge $(u', v') \in G$, in which case it is deleted, otherwise it is
created. Finally, the edges adjacent to $u$ and $v$ are swapped -- see
Fig.~\ref{elc2}. ELC on $G$ extends easily to ELC on $\TG$ when $H$ is
in standard form \cite{knudsen08}. Given a bipartite graph with
bipartition $(\mathcal{V}, \mathcal{U})$, we then have a one-to-one
mapping to a Tanner graph, with check nodes from the set $\mathcal{U}$
and bit nodes from $\mathcal{V} \cup \mathcal{U}$.
Fig. \ref{pivot} shows an example, where the bipartition is fixed
according to the sets $\mathcal{V}$ and $\mathcal{U}$. In
Fig. \ref{gb}, the left and right nodes correspond to $\mathcal{V}$
and $\mathcal{U}$, respectively, for the simple graph $G$. $\TG$ may
be obtained by replacing grey nodes by a check node singly connected
to a bit node, as illustrated in Fig.~\ref{ga}. Figs.~\ref{gb} and
\ref{ge} show an example of ELC on the edge $(0, 5)$. Although the
bipartition changes (edges adjacent to 0 and 5 are swapped),
Figs.~\ref{ga} and \ref{gd} show how the map to Tanner graphs, in
fact, preserves the code.

By complementing the edges of a local neighborhood of $\TG$, ELC has
the effect of row additions on $H$. The complexity of ELC on $(u,v)$
is $\mathcal{O}(|\mathcal{N}_u||\mathcal{N}_v|)$. The set of
vertex-labeled graphs generated by ELC on $\TG$ (or, equivalently,
$G$) is here called the \textit{ELC-orbit} of $\mathcal{C}$. Each
information set for $\mathcal{C}$ corresponds to a unique graph in the
ELC-orbit \cite{ELC_danielsen}. Note that this is a code property,
which, as such, is independent of the initial parity-check matrix,
$H$. The set of structurally distinct (unlabeled) graphs generated by
ELC is here called the \textit{s-orbit} of $\mathcal{C}$, and is a
subset of the ELC-orbit. Graphs are structurally distinct (i.e.,
non-isomorphic) if the corresponding parity-check matrices are not row
or column permutations of each other. Each structure in the s-orbit
has a set of $|\Autc|$ isomorphic graphs, comprising an
\textit{iso-orbit} \cite{knudsen08}. In the following, we will refer
to ELC directly on $\TG$, keeping Fig.~\ref{pivot} in mind.

\section{Decoding Algorithms}

\subsection{SPA}
The SPA is an inherently local algorithm on $\TG$, where the global
problem of decoding is partitioned into a system of simpler
subproblems \cite{kschiFact}. Each node and its adjacent edges can be
considered as a small constituent code, and essentially performs
maximum-likelihood decoding (MLD) based on local information. The key
to a successful decoder lies in this partitioning--how these
constituent codes are interconnected. The summed information contained
in a bit node, $v_i$, is the \textit{a posteriori} probability (APP),
$\hat{x}_i$, at codeword position $i$. The vector $\mathbf{\hat{x}}$
constitutes a tentative decoding of the received channel
vector, $\mathbf{y}$. The decoder input is the log-likelihood
ratio (LLR) vector $\mathbf{L} = (2/\sigma^2)\mathbf{y}$, where
$\sigma$ is the channel noise standard deviation on an additive white
Gaussian noise (AWGN) channel. Subtracting the input from the APP
leaves the extrinsic information, $\hat{x}_i - L_i$, which is
produced by the decoder.
The message on the edge from node $v$ to $u$, in the direction of $u$,
$\mu_{v \rightarrow u}$, is computed according to the SPA rule on node
$v$. The SPA computation of all check nodes, followed by all bit
nodes, is referred to as one \textit{flooding} iteration.

Classical codes, for which strong code properties are known, are
typically not very suitable for iterative decoding mainly due to the
high weight of their parity-check matrices, which gives many short
cycles in the corresponding Tanner graphs.

\subsection{Diversity Decoding}
A few recent proposals in the literature have attempted to enhance
iterative decoding by dynamically modifying $\TG$ during decoding, so
as to achieve diversity and avoid fixed points (local optima) in the
SPA convergence process. Efforts to improve decoding may, roughly, be
divided into two categories. The first approach is to employ several
structurally distinct matrices, and use these in a parallel, or
sequential, fashion \cite{jia06,hehn}. These matrices may be either
preprocessed, or found dynamically by changing the graph during
decoding. However, this incurs an overhead either in terms of memory
(keeping a list of matrices, as well as state data), or complexity
(adapting the matrix, e.g., by Gaussian elimination \cite{jia06}). The
other approach is to choose a code with a non-trivial $\Autc$, such
that diversity may be achieved by permuting the code coordinates
\cite{halford_NEW,jia04,beery,knudsen08}.
\begin{algorithm}[t]
\caption{SPA-PD$(I_1,I_2,I_3,\alpha_0)$ \cite{halford_NEW}}
\label{RRD}
\begin{algorithmic}[1]
\STATE // Input: $(\mathbf{y}, H, \alpha_0, I_1, I_2, I_3)$.
\STATE // Output: $\Theta^{-1}(\hat{\mathbf{x}})$.
\vspace{0.2cm}
\STATE $\alpha \gets \alpha_0$.
\FOR{$I_3$ times}
\STATE $\mathbf{L} \gets (2/\sigma^2)\mathbf{y}$ and $\Theta \gets \pi_0$  // identity permutation.
\FOR{$I_2$ times}
\STATE $\mu_{v \rightarrow u} \gets L_v, \;\forall (u,v) \in \TG$.
\STATE Do $I_1$ flooding iterations, $\hat{\mathbf{x}} \gets \mathrm{SPA}(\TG)$.
\STATE Take the hard decision of $\hat{\mathbf{x}}$ into $\mathbf{c}$, \textbf{stop} if $\mathbf{c}H^T = \mathbf{0}$.
\STATE $L_i \gets (\hat{x}_i - L_i)\alpha +L_i, \;\; 0 \le i < n$.
\STATE Draw random permutation $\pi \in \Aut{C}$ \cite{celler}.
\STATE $\mathbf{L} \gets \pi(\mathbf{L})$ and $\Theta \gets \pi(\Theta)$.

\ENDFOR
\STATE $\alpha \gets \alpha_0 + (1 - \alpha_0)\frac{I_3}{I_3 -1}$.
\ENDFOR
\end{algorithmic}
\end{algorithm}
An example is SPA-PD, listed in Algorithm~\ref{RRD}, where $\Autc$ is
represented by a small set of generators, and uniformly sampled using
an algorithm due to Celler \textit{et al.} \cite{celler}. These
permutations tend to involve all, or most, of the code coordinates,
making it a global operation.  
Note that line~7 in Algorithm~\ref{RRD} is to compensate for the fact
that permutations are applied to $\mathbf{L}$ in line~12, rather than
to the columns of $H$, after which the messages on the edges no longer
`point to' their intended recipients. This is yet another global
stage. The extrinsic information is damped by a coefficient $\alpha$,
$0 < \alpha \le 1$, in line~10 before being used to re-initialize the
decoder. Each time $\alpha$ is incremented, the decoder re-starts from
the channel vector, $\mathbf{y}$.

\subsection{SPA-ELC}
\label{spaelc}
Our proposed local algorithm is a two-stage iterative decoder,
interleaving the SPA with random ELC operations. We call this SPA-ELC,
and say that it realizes a local diversity decoding of the received
codeword. Our algorithm is listed in Algorithm~\ref{SPAELC}. Both
SPA-PD and SPA-ELC perform a maximum of $T \triangleq I_1I_2I_3$
iterations.
SPA update rules ensure that extrinsic information remains summed in
bit nodes, such that an edge may be removed from $\TG$ without loss of
information. New edges, $(u',v')$, should be initialized according to
line~13 in Algorithm~\ref{SPAELC}. Although neutral (i.e., LLR 0)
messages will always be consistent with the convergence process, our
experiments clearly indicate that this has the effect of `diluting'
the information, resulting in an increased decoding time and worse
error rate performance.

\begin{algorithm}[t]
\caption{SPA-ELC$(p, I_1, I_2, I_3, \alpha_0)$}
\label{SPAELC}
\begin{algorithmic}[1]
\STATE // Input: $(\mathbf{y}, H, \alpha_0, I_1, I_2, I_3, p)$.
\STATE // Output: $\hat{\mathbf{x}}$.
\vspace{0.2cm}
\STATE $\alpha \gets \alpha_0$.
\FOR{$I_3$ times}
\STATE $\mathbf{L} \gets (2/\sigma^2)\mathbf{y}$.
\STATE $\mu_{v \rightarrow u} \gets L_v, \;\forall (u,v) \in \TG$.
\FOR{$I_2$ times}
\STATE Do $I_1$ flooding iterations, $\hat{\mathbf{x}} \gets \mathrm{SPA}(\TG)$.
\STATE Take the hard decision of $\hat{\mathbf{x}}$ into $\mathbf{c}$, \textbf{stop} if $\mathbf{c}H^T = \mathbf{0}$.
\FOR {$p$ times}
\STATE Select random edge $e=(u,v) \in \TG$. 
\STATE $\TG \gets \mathrm{ELC}(\TG,e)$.
\STATE $\mu_{v' \rightarrow u'} \gets (\hat{x}_{v'} - L_{v'}) \alpha +
L_{v'}$,\\ $\forall (u',v') \in \TG, \; u' \in \mathcal{N}_v^u$, $v' \in \mathcal{N}_u^v$.
\ENDFOR
\ENDFOR
\STATE $\alpha \gets \alpha_0 + (1 - \alpha_0)\frac{I_3}{I_3 -1}$.
\ENDFOR
\end{algorithmic}
\end{algorithm}

The simple SPA-ELC decoder requires no preprocessing or any complex
heuristic or rule to decide when or where to apply ELC. As ELC
generates the s-orbit of $\C$, as well as the iso-orbit of each
structure, diversity of structure can be achieved even for random
codes, for which $|\Autc|$ is likely to be 1 while the size of the
s-orbit is generally very large. However, going outside the iso-orbit
means that we change the properties of $H$, most importantly in terms
of density and number of short cycles. Ideally, the SPA-ELC decoder
operates on a set of structurally distinct parity-check matrices,
which are all of minimum weight. With the exception of codes with very
strong structure, such as the extended Hamming code, the ELC-orbit of
a code will contain structures of weight greater than the
minimum. SPA-ELC should take measures against the negative impact of
increased weight. In this paper, we adapt the technique of damping to
our graph-local perspective. Damping with the standard SPA, where
$\TG$ is fixed, does not
work, 
so we only want to damp the parts of the graph which change. As
opposed to SPA-PD, only a subgraph of $\TG$ is affected by ELC, so we
restrict damping to new edges in line 13.
Note that SPA-ELC simplifies to a version without damping, denoted by
SPA-ELC$(p,I_1,T)$, when $\alpha_0 = 1$, $I_2 = T/I_1$, and $I_3 = 1$.
This is, simply, flooding iterations interspersed with random ELC
operations, where new edges are initialized with the adjacent APP
(line~13).

Currently, the SPA stopping criterion (i.e., the parameters used to
flag when decoding should stop) is still implemented globally. However, 
a reasonable local solution would be to remove the syndrome check
($\mathbf{c}H^T = \mathbf{0}$) from the stopping criterion, and simply
stop after $\hat{T}$ SPA-ELC iterations, where $\hat{T}$ can be
empirically determined. However, this has obvious implications for
complexity and latency. In some scenarios a stopping criterion can be
dispensed with anyway--for instance when using the decoder as some
form of distributed process controller, or for a pipelined
implementation in which the iterations are rolled out.

\section{Results}

We have compared SPA-ELC against standard SPA, and SPA-PD. Extended
quadratic residue (EQR) codes were chosen for the comparison, mainly
due to the fact that for some of these codes, $\Autc$ can be generated
by $3$ generators \cite{macw_sloane}. In fact, our experiments have
shown that EQR codes have Tanner graphs well-suited to SPA-ELC, at
least for short blocklengths. The codes considered have parameters
$[24, 12, 8]$ (the extended Golay code), $[48, 24, 12]$ (EQR48), and
$[104, 52, 20]$ (EQR104). Parity-check matrices for the codes were
preprocessed by heuristics to minimize the weight and the number of
$4$-cycles. The results are listed in Table~\ref{codes}, where columns
marked `W' and `C' show the weight and the number of $4$-cycles,
respectively. Columns marked `Initial' show the weight and the number
of $4$-cycles of the initial Tanner graph constructions. `Reduced' and
`Reduced IP' refer to optimized Tanner graphs, where the latter is
restricted to Tanner graphs in standard form. Entries marked by an
asterisk correspond to minimum weight parity-check matrices.

\begin{table}[t]
\caption{Optimization of codes used in simulations}
\vspace{-0.5cm}
\begin{tabular}[t]{l|r|r|r|r|r|r}
  \hline
  & \multicolumn{2}{c|}{Initial} & \multicolumn{2}{c|}{Reduced} & \multicolumn{2}{c}{Reduced IP} \\
  & \multicolumn{1}{c|}{W} & \multicolumn{1}{c|}{C} & \multicolumn{1}{c|}{W} & \multicolumn{1}{c|}{C} 
& \multicolumn{1}{c|}{W} & \multicolumn{1}{c}{C} \\
 \hline
  $[24, 12, 8]$ & * 96 & 366 & * 96 & 147   & * 96 & 366 \\
  \hline
  $[48, 24, 12]$ & 320 & 4936 & * 288 & 897 & * 288 & 2672 \\
  \hline
  $[104, 52, 20]$ & 1344 & 89138 & 1112 & 16946 & 1172 & 49839\\
  \hline
\end{tabular}
\label{codes}
\end{table}

\begin{figure}[t]
\centering
\subfigure[FER performance]{\label{fer_golay}\includegraphics[scale=0.35,angle=-90]
{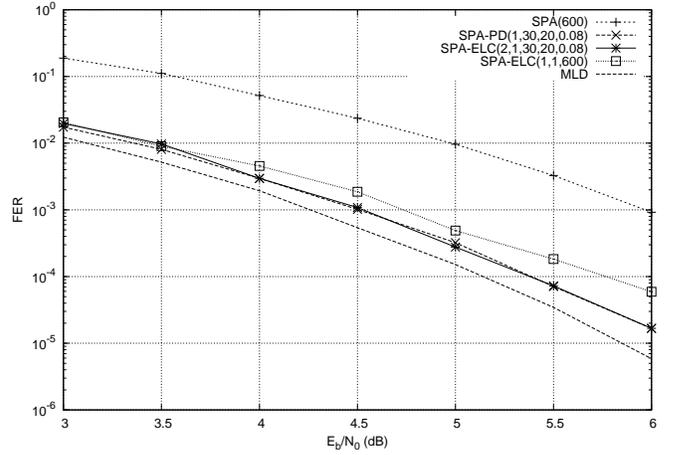}}
\subfigure[Average number of SPA messages]{\label{comp_golay}\includegraphics[scale=0.35,angle=-90]
{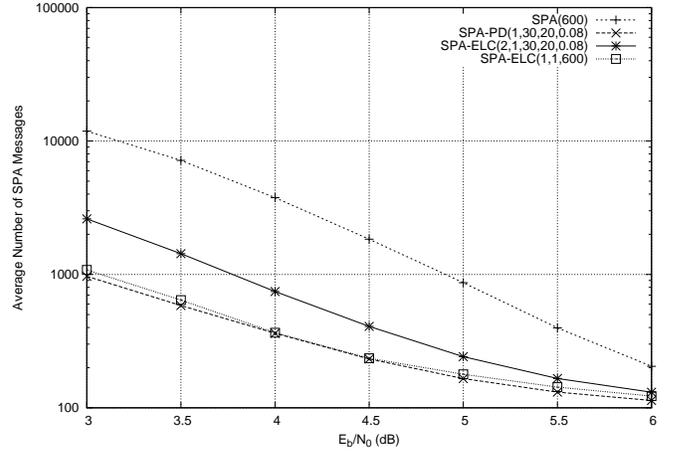}}
\caption{$[24, 12, 8]$ extended Golay code}
\label{beery_golay}
\end{figure}

\begin{figure}[t]
\centering
\subfigure[FER performance]{\label{fer_QR_48}\includegraphics[scale=0.35,angle=-90]
{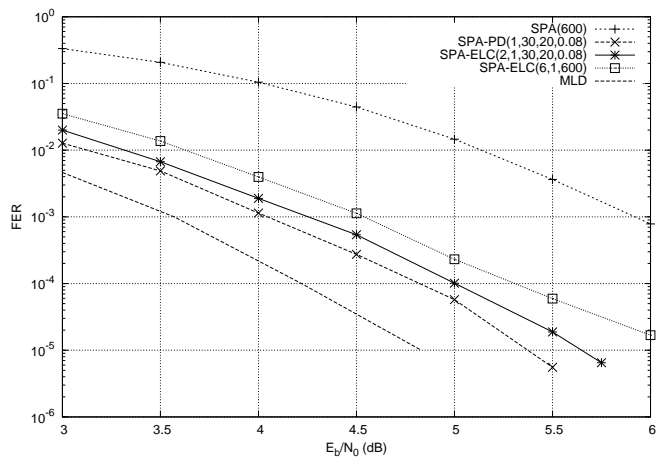}}
\subfigure[Average number of SPA messages]{\label{comp_QR_48}\includegraphics[scale=0.35,angle=-90]
{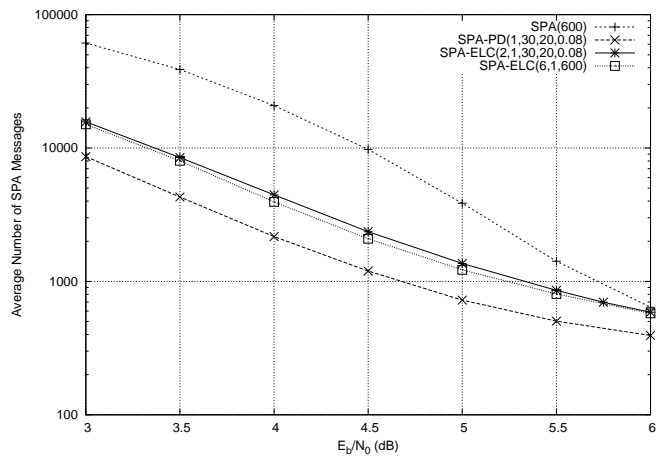}}
\caption{$[48, 24, 12]$ EQR48 code}
\label{QR48}
\end{figure}

\begin{figure}[t]
\centering
\subfigure[FER performance]{\label{fer_QR_104}\includegraphics[scale=0.35,angle=-90]
{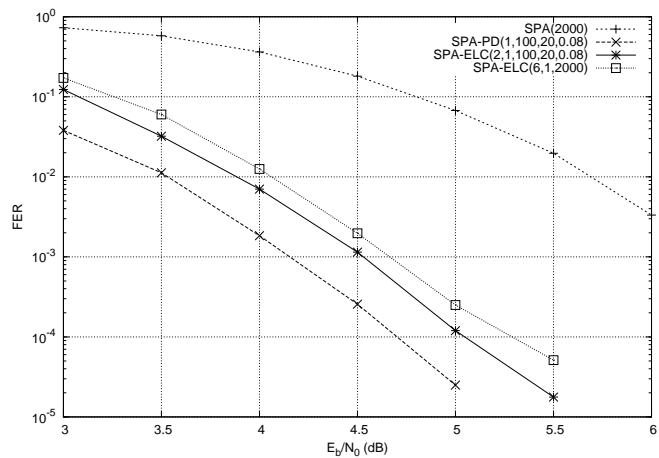}}
\subfigure[Average number of SPA messages]{\label{comp_QR_104}\includegraphics[scale=0.35,angle=-90]
{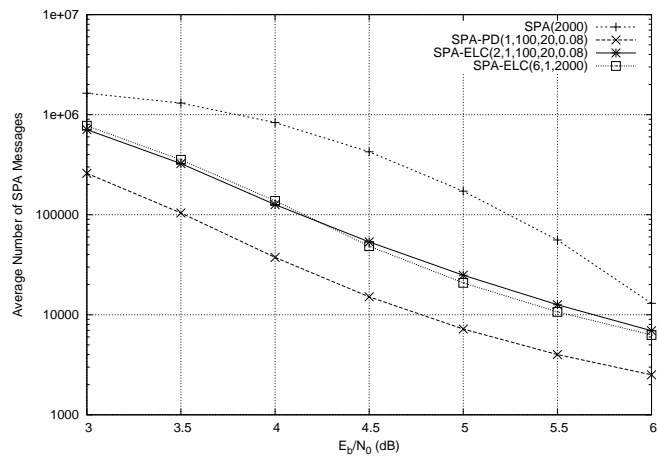}}
\caption{$[104, 52, 20]$ EQR104 code}
\label{QR104}
\end{figure}

In Figs.~\ref{beery_golay}-\ref{QR104}, we show the frame error rate
(FER) performance and the average number of SPA messages of SPA,
SPA-PD, and SPA-ELC for the extended Golay code, the EQR48 code, and
the EQR104 code, respectively, on the AWGN channel versus the
signal-to-noise ratio, $E_b/N_0$.

The specific parameters used are indicated in the figure legends. For
the extended Golay code and the EQR48 code, we set a maximum at $T =
600$ iterations, which we increased to $T=2000$ to accommodate the
larger EQR104 code. For SPA-ELC we have also included results without
damping. Since SPA-ELC changes the weight of $\TG$, we can not compare
complexity by simply counting iterations. Since the complexity of one
ELC operation is much smaller than the complexity of a SPA iteration,
the total number of SPA messages may serve as a common measure for the
complexity of the decoders. We have no initial syndrome check, so the
number of iterations approaches $1$ at high $E_b/N_0$. In the same
way, the complexity approaches the average weight of the matrices
encountered during decoding. Each FER point was simulated until at
least $100$ frame errors were observed.

From the figures, we observe that the SPA-ELC decoder outperforms
standard SPA decoding, both in terms of FER and decoding
complexity. The extended Golay code is a perfect example for
demonstrating the benefits of SPA-ELC. The s-orbit of this code
contains only two structures, where one is of minimum weight (weight
$96$) and the other only slightly more dense (weight $102$), while the
iso-orbit of the code is very large. Thus, we can extend SPA-PD with
multiple Tanner graphs (two structures) while keeping the density
low. Not surprisingly, SPA-ELC achieves the FER performance of SPA-PD,
albeit with some complexity penalty.  Note that the simple SPA-ELC
decoder, without damping, approaches closely the complexity of SPA-PD
at the cost of a slight loss in FER. For the larger codes, the sizes
of the s-orbits are very large, and many structures are less suited
for SPA-PD. Still, the same tradeoff between FER performance and
complexity holds, based on whether or not we use damping.  For the
EQR48 code, we have observed a rich subset of the s-orbit containing
minimum weight structures (weight $288$). The optimum value of $p$
(see line~10 in Algorithm~\ref{SPAELC}) was determined empirically.


\section{Conclusion and Future Work}

We have described a local diversity decoder, based on the SPA and the
ELC operation. The SPA-ELC algorithm outperforms the standard SPA both
in terms of error rate performance and complexity, and compares well
against SPA-PD, despite the fact that SPA-PD uses global
operations. Ongoing efforts are devoted to further improvements, and
include; selective application of ELC, rather than random; devise
techniques such that diversity may be restricted to sparse structures
in the s-orbit; identify a code construction suited to SPA-ELC, for
which the s-orbit contains several desirable structures even for large
blocklengths.

\section*{Acknowledgment}

The authors wish to thank Alban Goupil for providing the MLD curve for the EQR48 code.



\end{document}